\documentclass{article}

\usepackage{amsmath}
\usepackage{graphicx}
\usepackage{subcaption}
\usepackage{setspace}
\usepackage{multirow}
\usepackage{caption}
\usepackage[square]{natbib}
\usepackage{setspace}
\usepackage{lineno}
\usepackage{hyperref}
\usepackage[title]{appendix}

\begin{document}
\bibliographystyle{plainnat}

\title{Anti-diffusive, non-oscillatory central difference scheme (adNOC)
suitable for highly nonlinear advection-dominated problems.}
\author{
Haseeb Zia 
\thanks{
Dept. of Earth and Environmental Sciences,
Univ. of Geneva,
13 Rue des Maraichers,
1205 Geneva, Switzerland. 
E-mail: Haseeb.Zia@unige.ch}
\and Guy Simpson 
\thanks{
Dept. of Earth and Environmental Sciences,
Univ. of Geneva,
13 Rue des Maraichers,
1205 Geneva, Switzerland. 
E-mail: Guy.Simpson@unige.ch}
}

\maketitle

\begin{abstract}
Explicit non-oscillatory central difference schemes become excessively
diffusive when applied to highly nonlinear advection problems where small time
steps are necessary to maintain stability. Here, we present a correction to
reduce such numerical dissipation for this class of problems. The correction is
obtained by selecting the appropriate finite difference approximations for
calculating the slopes utilized to reconstruct the solution from the cell
averages. The anti-diffusive central scheme does not require any
knowledge of the eigenstructure and is fully central.
The proposed correction is applied to the widely used Nessyahu-Tadmor scheme to
demonstrate the utility of the correction. The stability of the corrected scheme is discussed and the condition for the scheme to become TVD (total variation
diminishing) is presented. The corrected scheme is finally tested with a number
of test cases and the results are compared with analytical solutions and
published results showing the ability of the corrected scheme to
effectively resolve sharp discontinuities.
\end{abstract}

\singlespacing

\section{Introduction}
\label{sec:Introduction}

Advection dominated problems described by hyperbolic conservation laws are of
major importance in many areas of science and engineering. Because these
problems are often highly nonlinear, for example due to the presence of
propagating shock waves, analytical solutions are often complicated or  not available. Therefore, the development of accurate and robust numerical methods to solve hyperbolic conservation laws  is a domain of intense ongoing research.

Many of the schemes developed to solve hyperbolic conservation laws have their
origin in Godunov's method \cite{godunov1959difference}. These schemes are
essentially upwind conservative finite volume methods where numerical fluxes are
computed at cell interfaces based on local Riemann problems (e.g., see
\cite{toro2009riemann,leveque1992numerical} for more details). While these
methods can be very accurate, they may suffer from several potential drawbacks.
First, for some very nonlinear problems, the wave structure of the governing
equations may  not be  known, which makes it difficult to accurately compute
interface fluxes (e.g. see \cite{caleffi2007high,balbas2004non}). Second,
conservation laws with source terms are often treated with operator splitting, which can in some cases lead to significant
numerical errors (e.g. see
\cite{engquist1999multiphase,jiang1998nonoscillatory,balbas2004non}).

An alternative to the Godunov approach are central differencing finite volume
schemes, due largely to the work of Nessyahu and Tadmor
\cite{NessyahuTadmor1990}. Central differencing schemes are based on the
Lax-Friedrichs (LxF) scheme that is normally modified to include higher order
accuracy
\cite{liu1998third,huynh1995piecewise,bianco1999high,qiu2002construction} and
several dimensions \cite{arminjon1995two,jiang1998nonoscillatory,katsaounis1999modified,levy2002fourth,balbas2009non}.
These schemes have become known as non-oscillatory central differencing (NOC) methods. These
schemes involve no Riemann problems (and therefore require no knowledge of the
eigenstructure of the governing equations) and necessitate no operator
splitting. Thus, they are relatively simple and especially suitable for the
solution of highly nonlinear hyperbolic conservation laws  involving stiff
source terms (e.g., see application of central schemes in
\cite{tai2002shock,bryson2005high,anile2001assessment,kurganov2011semi,chertock2014central}).
A major factor limiting the utility of the central differencing schemes has  been that they introduce excessive numerical diffusion \cite{kurganov2000new,huynh2003analysis,abreu2009central,kurganov2007reduction,siviglia2013numerical,canestrelli2012restoration}.
In the original  central differencing schemes, this diffusivity was shown to be 
of order $O((\Delta x)^{2r}/\Delta t)$ \cite{kurganov2000new}, where $r$ is the
order of the scheme, showing that the numerical diffusion becomes more important
as the time step is reduced. Thus, for very nonlinear systems where small time
steps are of paramount importance to ensure stability, the solution may be
partially or completely destroyed by artificial diffusion \cite{kurganov2000new,kurganov2007reduction,stecca2012finite}.  Less diffusive modified schemes
utilizing partial knowledge of
eigenstructure have been proposed \cite{kurganov2000new,kurganov2002central},
but excessive diffusivity remains a limitation to  central differencing schemes
when small time steps are necessary.

In this article we present a simple anti-diffusion correction to the classic
NOC scheme \cite{NessyahuTadmor1990} in an effort
to reduce numerical dissipation when small time steps  must be used. As the
original scheme, our method utilizes  a staggered grid where the solution is
approximated by reconstructing piece-wise polynomials within the cells from the
evolving cell averages. The staggered approach enables the central scheme to
have smooth cell interfaces, which makes evaluation of numerical fluxes
particularly straight forward. Unlike previously proposed modifications
\cite{kurganov2000new,kurganov2002central} which require partial knowledge of
the eigenstructure, this scheme does not involve the
solution of Riemann problems and does not require any knowledge of the
eigenstructure of the governing system.  Here we demonstrate the ability of the
anti-diffusive, non-oscillatory central difference scheme (adNOC) to solve the shallow water equations coupled to substrate erosion and sedimentation.

The article is structured as follows. In section \ref{sec:Central schemes}, a brief description of Nessyahu-Tadmor
central scheme is presented. Section \ref{sec:Anti-diffusion slopes} describes how the diffusion can be eliminated by
using appropriate finite difference approximations. Section
\ref{sec:Stability} discusses the stability of the corrected scheme and finally, test cases and results are
presented in section \ref{sec:Test cases}.

\section{Central schemes}
\label{sec:Central schemes}
Consider the following scalar hyperbolic conservation law
\begin{equation}\label{eq:1}
\frac{\partial u}{\partial t}+\frac{\partial f(u)}{\partial 
x}=s(u),
\end{equation}
where $u$ is the conserved quantity, $f$ is the flux and $s$ is 
the source term,
both functions of $u$. To explain the centred approach, we will 
use the
Nessyahu-Tadmor scheme \cite{NessyahuTadmor1990}, the most widely used second
order method as the
standard central scheme. The development of this section follows closely that
presented by \cite{pudasaini2007avalanche}.
The method is a high-order extension of the Lax-Friedrichs solver 
which operates in predictor corrector fashion. The predictor
step involves evaluation of first order approximations at half 
time steps. The second
order solution is then realized in the corrector step which 
utilizes the
calculations from the predictor step to evaluate the solution on
the staggered cells. Below, we present a description of the 
procedure involved
in evaluation of the solution using the NOC scheme.

We begin by dividing spatial domain into cells.
Let $C_{j}$ denote the cell that covers the region $
|x-x_{j}|\le\frac{\text{\ensuremath{\Delta}}x}{2}$ where $\Delta 
x$ is the constant  grid
spacing. For the development below, we note that  the  cell  
$C_{j+1/2}$
consists of the overlap between the two adjacent cells $C_{j}$ and 
$C_{j+1}$.
Let $\overline{u}{}_{j}^{n}$ denote the cell average over the cell 
at time
$t^{n}$. The solution can be reconstructed in space linearly over 
the cell from
the average by:

\begin{equation}\label{eq:2}
u_{j}(x,t^{n})=\overline{u}{}_{j}^{n}+\sigma_{j}^{x}(x-
x_{j}),\,\hspace{15pt}
(x) \; \epsilon \; C_{j},
\end{equation}
where $\sigma^{x}$ is the discrete spatial slope of
the solution $u$  evaluated at time $t^{n}$  (i.e., $\sigma^{x} = \partial u / \partial x$). This reconstruction
is used to achieve second-order accuracy in space. Second-order
temporal accuracy is achieved by using a predictor-corrector 
procedure in which
the solution is first evaluated at the half time step in the 
predictor step. Linear reconstruction is also used for 
reconstruction in time:

\begin{equation}\label{eq:3}
\overline{u}_{j}^{n+1/2}=\overline{u}_{j}^{n}+\frac{\Delta
t}{2}\left(\frac{\partial u}{\partial t}\right)^{n},
\end{equation}
where $\left(\frac{\partial u}{\partial t}\right)^{n}$ is 
calculated using the
conservation law, i.e. Eq. (\ref{eq:1}):

\begin{equation}\label{eq:4}
\left(\frac{\partial u}{\partial t}\right)^{n}=-
\left(\frac{\partial
f(u)}{\partial x}\right)^{n}+s(\overline{u}^{n}).
\end{equation}
Thus the predictor step is given by:
\begin{equation}\label{eq:5}
\overline{u}_{j}^{n+1/2}=\overline{u}_{j}^{n}-\frac{\Delta
t}{2}(\sigma^{f})_{j}^{n}+\frac{\Delta t}
{2}s(\overline{u}_{j}^{n}),
\end{equation}
where $\sigma^{f}$ is the discrete spatial slope of the
flux. This first order solution evaluated in the predictor step will be 
used later
for second order evaluation of the solution.

\begin{figure}
\centering
\includegraphics{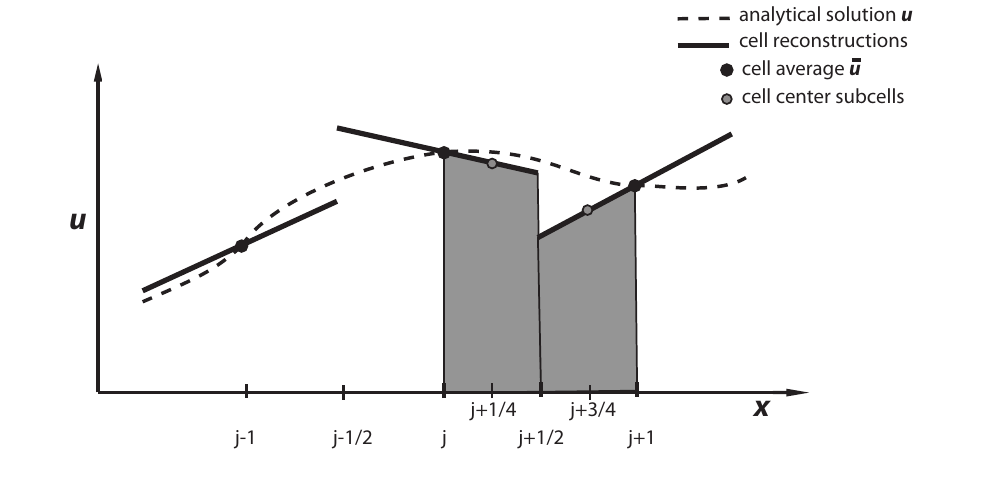}

\caption{Cell reconstructions and quadrature evaluations.}

\label{fig:quadrature}
\end{figure}

To calculate the second order solution for the conservation law 
given by Eq.
(\ref{eq:1}), we start by integrating it over the cell 
$C_{j+\frac{1}{2}}$
and time period {[}$t^{n},t^{n+1}${]}.
\[
\int_{x_{j}}^{x_{j+1}}u(x,t^{n+1})dx=\int_{x_{j}}^{x_{j+1}}u(x,t^{
n})dx
-\int_{t^{n}}^{t^{n+1}}\{f(x_{j+1},t)-f(x_{j},t)\}dt
\]

\begin{equation}\label{eq:6}
+\int_{x_{j}}^{x_{j+1}}\int_{t^{n}}^{t^{n+1}}s(x,t)dtdx.
\end{equation}
 Notice that the integral is evaluated at
the staggered cell. This allows the evaluation of fluxes at the 
centre of the
cells from previous time step where the reconstructions are 
smooth, see Fig.
\ref{fig:quadrature}.
This is different from upwind schemes where the quadrature is 
evaluated in the
smooth region and the flux is evaluated at the cell interfaces 
where the
piece-wise reconstructions are discontinuous. Note that the slope 
used
for reconstruction in Fig. \ref{fig:quadrature} are standard forward 
difference
approximations. A different finite difference formulation would 
significantly
change the reconstructions. The impact of different choice of 
finite difference
formulation will be discussed in more detail in the next section. 
Evaluating
integrals using the second order accurate midpoint rule gives:
\[
\Delta x \; \overline{u}_{j+1/2}^{n+1}=\frac{\Delta
x}{2}(\overline{u}_{j+1/4}^{n}+\overline{u}_{j+3/4}^{n})-\Delta
t(f_{j+1}^{n+1/2}-f_{j}^{n+1/2})
\]

\begin{equation}\label{eq:7}
+\frac{\Delta t\Delta
x}{2}(s_{j+1/4}^{n+1/2}+s_{j+3/4}^{n+1/2}).
\end{equation}
The first integral on right hand side of Eq. 
(\ref{eq:6})  consists of  two
parts i.e. quadrature of two sub-cells with $j+1/4$ and $j+3/4$ as 
cell centres.
This is performed with discontinuous piece-wise cell 
reconstructions, as shown
with grey regions in Fig. \ref{fig:quadrature}. Dividing Eq. 
(\ref{eq:7}) by
$\Delta x$ leads to

\begin{equation}\label{eq:8}
\overline{u}_{j+1/2}^{n+1}=\frac{1}{2}
(\overline{u}_{j+1/4}^{n}+\overline{u}_{j+3/4}^{n})-\frac{\Delta
t}{\Delta x}(f_{j+1}^{n+1/2}-f_{j}^{n+1/2})+\frac{\Delta
t}{2}(s_{j+1/4}^{n+1/2}+s_{j+3/4}^{n+1/2}),
\end{equation}
\begin{doublespace}
\end{doublespace}
where $\overline{u}_{j+1/4}^{n}$ and $\overline{u}_{j+3/4}^{n}$ 
are approximated
by reconstructions
\begin{equation}\label{eq:9}
\overline{u}_{j+1/4}^{n}=\overline{u}_{j}^{n}+\frac{\Delta
x}
{4}\sigma_{j}^{n},\;\;\overline{u}_{j+3/4}^{n}=\overline{u}_{j+1}^
{n}-\frac{\Delta
x}{4}\sigma_{j+1}^{n}.
\end{equation}
The integral of fluxes in Eq. (\ref{eq:6})  is approximated using the half time
step solution
evaluated in the predictor step, see Eq. (\ref{eq:5}).
\begin{equation}\label{eq:10}
f_{j}^{n+1/2}=f(\overline{u}_{j}^{n+1/2}),\;\;
f_{j+1}^{n+1/2}=f(\overline{u}_{j+1}^{n+1/2}).
\end{equation}
Similarly, the integral of source terms is approximated by
\[
s_{j+1/4}^{n+1/2}=s(\overline{u}_{j+1/4}^{n+1/2}),\;\;
s_{j+3/4}^{n+1/2}=s(\overline{u}_{j+3/4}^{n+1/2}),
\]
where 
\[
\overline{u}_{j+1/4}^{n+1/2}=\overline{u}_{j}^{n+1/2}+\frac{\Delta
x}
{4}\sigma_{j}^{n},\;\;\overline{u}_{j+3/4}^{n+1/2}=\overline{u}_{j
+1}^{n+1/2}-\frac{\Delta
x}{4}\sigma_{j+1}^{n}.
\]

It is important to remember that while the high order nature of 
the scheme assures
that shocks and discontinuities are captured, this comes at the 
expense of
spurious oscillations. To avoid these oscillations, the spatial 
slopes
utilized for reconstructions and the flux slopes should be
evaluated with limiters. The use of slope and flux limiters 
together with the
generalized CFL (Courant-Friedrichs-Lewy) condition satisfy the 
TVD (total
variation diminishing) condition, ensuring stability of the scheme. 
More on
stability and robustness of scheme is discussed in section \ref{sec:Stability}. 

\section{Anti-diffusion slopes}
\label{sec:Anti-diffusion slopes}
As mentioned in the introduction, classic NOC scheme becomes excessively diffusive when small time steps are 
used or when flux difference and source terms in Eq. (\ref{eq:8}) are relatively
small.
This may occur during near steady-state flow conditions when the flux is same
throughout  the domain or in passive stages when the flux is zero.
To understand this, consider the conservation law (Eq. \ref{eq:1}) without the
source term. In the case of steady or passive state, the second
integral on right hand side of Eq. (\ref{eq:6}) becomes zero. In this case, Eq. (\ref{eq:8}) for any cell $C_j$ can be
written as:
\[
\overline{u}_{j}^{n+1}=\frac{1}{2}(\overline{u}_{j-
1/4}^{n}+\overline{u}_{j+1/4}^{n})
\]
\begin{equation}\label{eq:11}
=\frac{1}{2}(\overline{u}_{j-
1/2}^{n}+\overline{u}_{j+1/2}^{n})+\frac{\Delta
x}{8}(\sigma_{j-1/2}^{n}-\sigma_{j+1/2}^{n}).
\end{equation}
Substituting 
\[
\overline{u}_{j-1/2}^{n}=\frac{1}{2}(\overline{u}_{j-1}^{n-
1}+\overline{u}_{j}^{n-1})+\frac{\Delta
x}{8}(\sigma_{j-1}^{n-1}-\sigma_{j}^{n-1})
\]
and
\[
\overline{u}_{j+1/2}^{n}=\frac{1}{2}(\overline{u}_{j}^{n-
1}+\overline{u}_{j+1}^{n-1})+\frac{\Delta
x}{8}(\sigma_{j}^{n-1}-\sigma_{j+1}^{n-1})
\]
into Eq. (\ref{eq:11}) results in
\[
\overline{u}_{j}^{n+1}=\overline{u}_{j}^{n-1}+\frac{1}{4}
(\overline{u}_{j-1}^{n-1}-2\overline{u}_{j}^{n-
1}+\overline{u}_{j+1}^{n-1})
\]
\begin{equation}\label{eq:12}
+\frac{\Delta
x}{8}(\sigma_{j-1}^{n-1}-\sigma_{j+1}^{n-1})+\frac{\Delta
x}{8}(\sigma_{j-1/2}^{n}-\sigma_{j+1/2}^{n}).
\end{equation}
The second term on the right hand side of Eq. (\ref{eq:12}) is the 
finite
difference approximation for a diffusive term (ie., $
\frac{1}{4}(\overline{u}_{j-1}^{n-1}-2\overline{u}_{j}^{n-
1}+\overline{u}_{j+1}^{n-1})\approx\frac{(\Delta
x)^{2}}{4}\frac{\partial^{2}u}{\partial x^{2}}$), which shows the origin of
numerical dissipation in this scheme.

\begin{figure}
\centering
\includegraphics{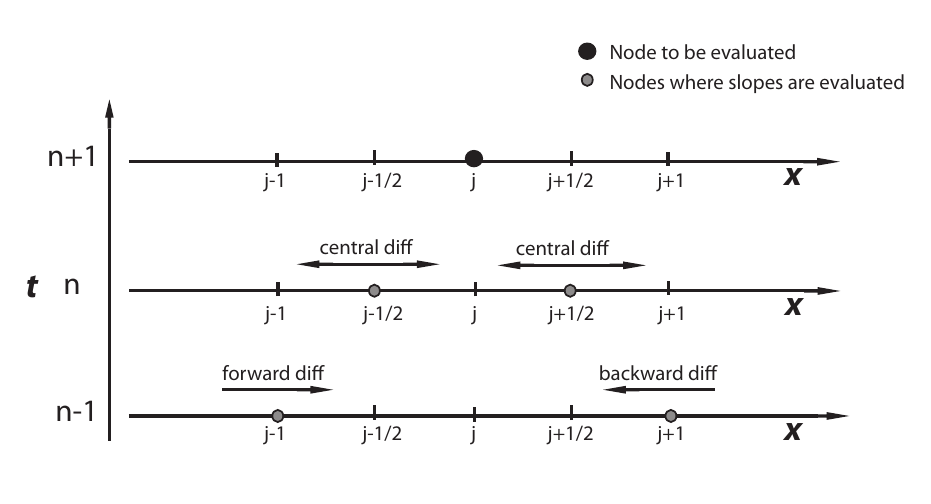}

\caption{Finite difference approximations to be utilized for anti-diffusive
scheme.}
\label{fig:approxs}
\end{figure}

This dissipation, however, can be mitigated or removed entirely 
by carefully choosing the finite difference approximations
of the slopes present in the equation. Indeed, finite 
difference approximations with the
difference direction towards the cell centre fulfil this 
requirement. This allows the local slopes to be used in
calculation, avoiding the use of slopes evaluated from adjacent 
cells which
are not relevant (see Fig. \ref{fig:approxs}). These slopes are 
given by:

\begin{equation}\label{eq:13}
\sigma_{j-1}^{n-1}=\sigma_{j-1/2}^{n}=\frac{\overline{u}_{j}^{n-
1}-\overline{u}_{j-1}^{n-1}}{\Delta x}
\end{equation}
and
\begin{equation}\label{eq:14}
\sigma_{j+1}^{n-1}=\sigma_{j+1/2}^{n}=\frac{\overline{u}_{j+1}^{n-
1}-\overline{u}_{j}^{n-1}}{\Delta
x}.
\end{equation}
Substituting (\ref{eq:13}) and (\ref{eq:14}) into Eq. (\ref{eq:12}) 
reduces it to
\[
\overline{u}_{j}^{n+1}=\overline{u}_{j}^{n-1},
\]
Thus,  the solution is exactly maintained in the  case of steady 
and
passive states without any smearing. It has to be noted that the slopes used
here are from the previous time step which means that the anti-diffusive scheme has a 3 level deep stencil in time.

\begin{doublespace}

\end{doublespace}

\section{Stability}
\label{sec:Stability}
The Nessyahu-Tadmor scheme is a Total-Variation-Diminishing method. 
The concept of
TVD (total variation diminishing), first introduced by Harten
\cite{harten1983high} has been utilized by many schemes to avoid 
spurious
oscillations when high resolution schemes are used. 
The total
variation of a conserved quantity, $\overline{u}^n$ at any time 
step $n$ is defined as:
\[
TV(\overline{u}^{n})=\sum_{j=0}^{N-1}(\overline{u}_{j+1}^{n}-
\overline{u}_{j}^{n}).
\]
Numerical oscillations 
increase the
\textit{total variation} and may render the scheme unstable. However, if the 
total variation
is ensured not to increase through the time evolution, the scheme 
is said
to be \textit{Total-Variation-Diminishing}. For any time step $n$, the TVD 
condition
is given as:
\[
TV(\overline{u}^{n+1})\leq TV(\overline{u}^{n})
\]

TVD methods are \textit{monotonicity preserving}, ensuring that 
spurious
oscillations do not arise near propagating discontinuities. 
Proof
of the  Nessyahu-Tadmor scheme being TVD can be seen in the paper 
introducing the
scheme \cite{NessyahuTadmor1990}. Next, we will see how the anti-diffusion
slopes presented in the previous section affect the stability of 
the scheme.
We will start with the lemma presented in 
\cite{NessyahuTadmor1990} which states
that the scheme is TVD if the following condition is held:
\begin{equation}\label{eq:15}
\lambda\mid\frac{\Delta
F_{j+1/2}}{\Delta\overline{u}_{j+1/2}}\mid\leq\frac{1}{2},
\end{equation}
where,
\begin{equation}\label{eq:16}
F_{j}=f(\overline{u}_{j}(t+\frac{\Delta
t}{2}))+\frac{1}{8\lambda}\overline{u}',\;\;\;\Delta
F_{j+1/2}=F_{j+1}-F_{j},
\end{equation}
and $\lambda$ is $\Delta t/\Delta x$. Substituting (\ref{eq:16}) 
into (\ref{eq:15}) gives:
\[
\lambda\mid\frac{\Delta F_{j+1/2}}
{\Delta\overline{u}_{j+1/2}}\mid\leq\lambda\mid\frac{f(\overline{u
}_{j+1}(t+\frac{\Delta t}{2}))-f(\overline{u}_{j}(t+\frac{\Delta 
t}{2}))}{\Delta\overline{u}_{j+1/2}}\mid+\frac{1}
{8}\mid\frac{\Delta\overline{u}'_{j+1/2}}
{\Delta\overline{u}_{j+1/2}}\mid
\]
\begin{equation}\label{eq:17}
\leq\lambda\mid\frac{f(\overline{u}_{j+1}(t+\frac{\Delta
t}{2}))-f(\overline{u}_{j}(t+\frac{\Delta t}{2}))}
{\overline{u}_{j+1}(t+\frac{\Delta t}{2})-\overline{u}_{j}
(t+\frac{\Delta t}{2})}\mid.\mid\frac{\overline{u}_{j+1}
(t+\frac{\Delta t}{2})-\overline{u}_{j}(t+\frac{\Delta t}{2})}
{\Delta\overline{u}_{j+1/2}}\mid+\frac{1}
{8}\mid\frac{\Delta\overline{u}'_{j+1/2}}
{\Delta\overline{u}_{j+1/2}}\mid
\end{equation}
We assume that the product $\lambda.\max_{j}\mid
a(\overline{u}_{j})\mid\leq\beta$ where $a(\overline{u}_j)$ is the
velocity at any cell centre $j$. This implies that the first term in 
the
inequality (\ref{eq:17}) is bounded by $\beta$:
\[
\lambda\mid\frac{f(\overline{u}_{j+1}(t+\frac{\Delta
t}{2}))-f(\overline{u}_{j}(t+\frac{\Delta t}{2}))}
{\overline{u}_{j+1}(t+\frac{\Delta t}{2})-\overline{u}_{j}
(t+\frac{\Delta t}{2})}\mid\leq\beta.
\]
Using the mid time step value from Eq. (\ref{eq:5}) without the source term leads to
\begin{equation}\label{eq:18}
\mid\frac{\overline{u}_{j+1}(t+\frac{\Delta
t}{2})-\overline{u}_{j}(t+\frac{\Delta
t}{2})}{\Delta\overline{u}_{j+1/2}}\mid\leq1+\frac{\lambda}
{2}\mid\frac{\Delta
f'_{j+1/2}}{\Delta\overline{u}_{j+1/2}}\mid.
\end{equation}
Slope limiters are a widely used tool to ensure stability of numerical
schemes (e.g., see \cite{NessyahuTadmor1990} \cite{sweby1984high}). To get an upper bound on the term $\mid\frac{\Delta
f'_{j+1/2}}{\Delta\overline{u}_{j+1/2}}\mid$, we assume that slopes are
limited by the \textit{minmod}
limiter given by,
\[
minmod\{r_1,r_2\}=\frac{1}{2}[sgn(r_1)+sgn(r_2)].Min(|r_1|,|r_2|).
\]
Since the  \textit{minmod} limiter ensures that the sign of consecutive slopes
$f'_{j+1}$ and $f'_{j}$ can not be different,
\begin{equation}\label{eq:19}
\mid\frac{\Delta f'_{j+1/2}}
{\Delta\overline{u}_{j+1/2}}\mid\leq\mid\frac{\mid f'_{j+1}\mid-
\mid f'_{j}\mid}{\Delta\overline{u}_{j+1/2}}\mid\leq 
max(\mid\frac{f'_{j+1}}
{\Delta\overline{u}_{j+1/2}}\mid,\mid\frac{f'_{j}}
{\Delta\overline{u}_{j+1/2}}\mid)\leq\frac{1}{\lambda}\beta
\end{equation}

\begin{figure}
		\centering
		\includegraphics[scale=0.6]{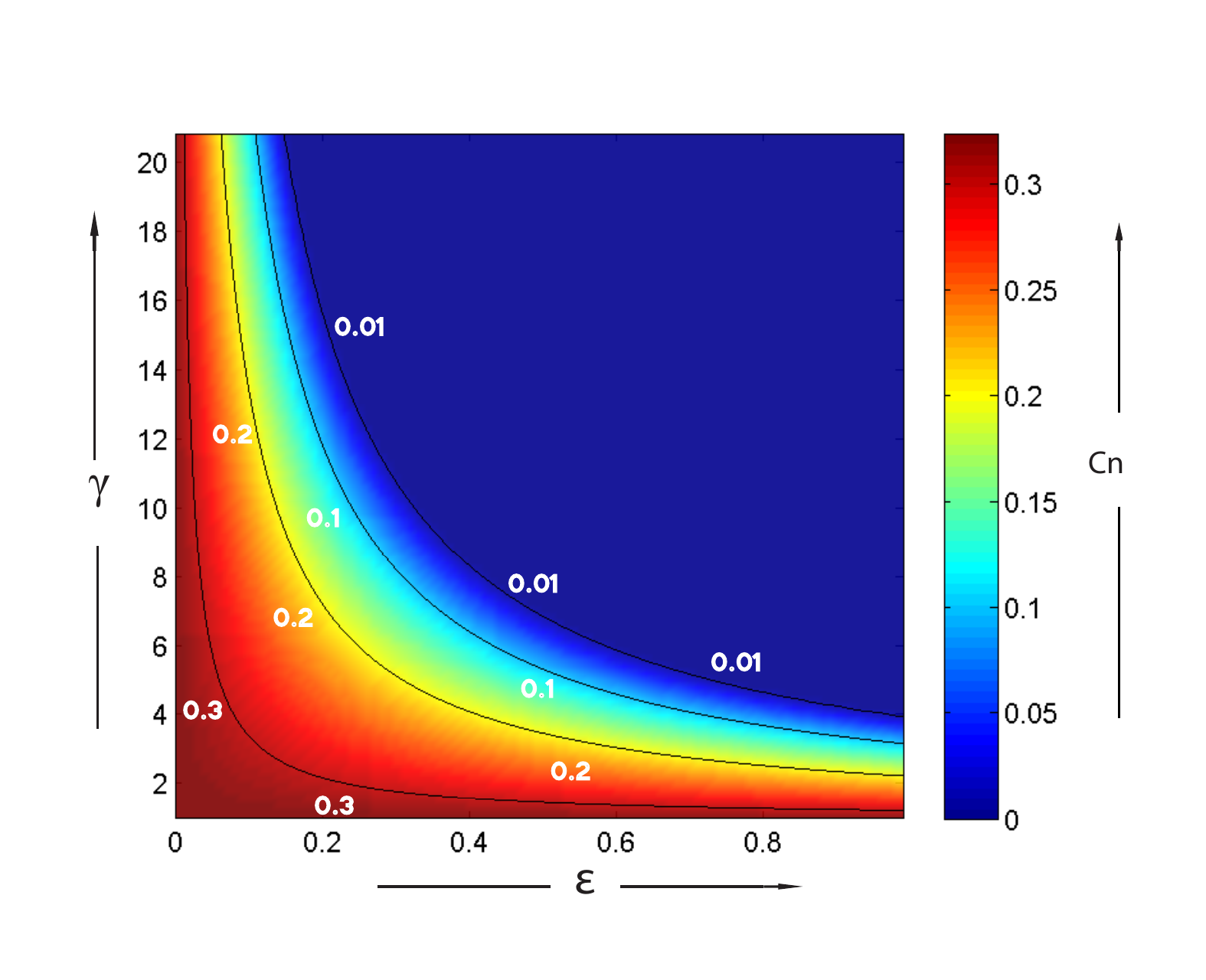}
		\caption{Maximum stable courant number for different combinations 
				of $\gamma$ and $\varepsilon$ evaluated using the inequality (\ref{eq:22}).}
		\label{fig:nd_param}
	\end{figure}

The third term in Eq. (\ref{eq:17}) can be decomposed into two
components, 
\[
\overline{u}'_{j}=(1-\varepsilon)\overline{u}_{j}^{'limited}+
(\varepsilon)\overline{u}_{j}^{'nd},
\]
where $\overline{u}_{j}^{'limited}$ is the limited difference (limited by the
\textit{minmod} limiter in this case), $\overline{u}_{j}^{'nd}$ 
is the
difference evaluated using the anti-diffusive finite difference 
approximations described in the
previous section and $\varepsilon$ is a factor signifying the strength of the
anti-diffusive slopes used. A value of 1 for $\varepsilon$ 
signifies that only anti-diffusive slopes are used in calculation of reconstructions while a
value of 0 signifies that the standard NOC scheme is used.
The term $\mid\frac{\Delta
\overline{u}'_{j+1/2}}{\Delta\overline{u}_{j+1/2}}\mid$ in the case of limited
difference is bounded by:
\begin{equation}\label{eq:20}
\mid\frac{\Delta\overline{u}'_{j+1/2}}
{\Delta\overline{u}_{j+1/2}}\mid\leq\mid\frac{\mid\overline{u}'_{j
+1}\mid-\mid\overline{u}'_{j}\mid}
{\Delta\overline{u}_{j+1/2}}\mid\leq 
max(\mid\frac{\overline{u}'_{j+1}}
{\Delta\overline{u}_{j+1/2}}\mid,\mid\frac{\overline{u}'_{j}}
{\Delta\overline{u}_{j+1/2}}\mid)\leq1.
\end{equation}
The upper bound for the anti-diffusive difference is given by $\gamma$, where
\begin{equation}\label{eq:21}
\gamma=max(\mid\frac{\overline{u}_{j}^{n}-\overline{u}_{j-1}^{n}}
{\overline{u}_{j+1}^{n}-
\overline{u}_{j}^{n}}\mid,\mid\frac{\overline{u}_{j}^{n-1}-
\overline{u}_{j-1}^{n-1}}{\overline{u}_{j+1}^{n}-
\overline{u}_{j}^{n}}\mid,\mid\frac{\overline{u}_{j+1}^{n-1}-
\overline{u}_{j}^{n-1}}{\overline{u}_{j+1}^{n}-
\overline{u}_{j}^{n}}\mid,1).
\end{equation}
Using the inequalities (\ref{eq:18}), (\ref{eq:19}), (\ref{eq:20}) 
and
(\ref{eq:21}), Eq. (\ref{eq:17}) is reduced to:
\begin{equation}\label{eq:22}
\beta(1+\frac{1}{2}\beta)+\frac{1}{8}((1-\varepsilon)+
(\varepsilon)\gamma)\leq\frac{1}{2}
\end{equation}
The inequality equation can be solved to get the upper bound $\beta$. This bound
$\beta$ is equivalent to the maximum stable courant
number ($Cn$) which is used to calculate the stable time step. Fig. 
\ref{fig:nd_param} shows
the maximum stable courant number for different combinations of 
$\gamma$ and $\varepsilon$. In the
case of coupled systems, where the time step is controlled by some 
other
process, the condition (\ref{eq:22}) can be used to determine the 
maximum stable ratio
$\varepsilon$ of the anti-diffusive slopes and the limited slopes. 
In practice, a
higher $\varepsilon$ can be used, depending on the systems being 
coupled. The
NOC scheme corrected with anti-diffusion slopes is given  by:
\[
\overline{u}_{j+1/2}^{n+1}=\frac{1}{2}
(\hat{u}_{j+1}^{n}+\hat{u}_{j}^{n})+\frac{\Delta x}{8}(1-
\varepsilon)(\sigma_{j}^{n}-\sigma_{j+1}^{n})-\frac{\varepsilon}
{4}(\overline{u}_{j+3/2}^{n-1}-2\overline{u}_{j+1/2}^{n-
1}+\overline{u}_{j-1/2}^{n-1})
\]
\begin{equation}\label{eq:23}
\;\;\;\;-\lambda(f_{j+1}^{n+1/2}-f_{j}^{n+1/2})+\frac{\Delta
t}{2}(s_{j+1/4}^{n+1/2}+s_{j+3/4}^{n+1/2}).
\end{equation}
where $\hat{u}$ is the cell average evaluated without the anti-diffusion
correction i.e. the third term in the right hand side of (\ref{eq:23}). The two
dimensional version of the anti-diffusive NOC scheme is presented in appendix
\ref{app:anti-diffusive central scheme in 2-dimensions}.

\begin{doublespace}
\end{doublespace}

\section{Test cases}
\label{sec:Test cases}
We have performed a number of simulations to test the success of the anti-diffusion correction in reducing numerical dissipation in the standard NOC scheme.
All of the cases presented here deal with the shallow water equations, either with or without substrate erosion and sedimentation.

\subsection{Shallow water Equations}
The shallow water equations are a depth 
averaged
reformulation of the Navier-Stokes equations that are widely used for
modelling of open surface hydraulics. The governing equations are:
\begin{equation}\label{eq:24}
\frac{\partial h}{\partial t}+\frac{\partial(hu)}{\partial 
x}+\frac{\partial(hv)}{\partial
y}=0,
\end{equation}
\begin{equation}\label{eq:25}
\frac{\partial(hu)}{\partial t}+\frac{\partial}{\partial
x}(hu^{2}+\frac{1}{2}gh)+\frac{\partial}{\partial y}(huv)=0,
\end{equation}
\begin{equation}\label{eq:26}
\frac{\partial(hv)}{\partial t}+\frac{\partial}{\partial x}
(huv)+\frac{\partial}{\partial
y}(hv^{2}+\frac{1}{2}gh^{2})=0,
\end{equation}
where $h$ is the water depth, $u$ and $v$ are the (depth-averaged)  velocities in the
$x$ and $y$
directions, respectively and $g$ is the gravitational acceleration. In this
form, bed slope, bed friction and substrate erosion/sedimentation are neglected.
The equations represent a set of hyperbolic conservative laws which can be solved with a variety of numerical methods. In the following test cases, we solve these equations
for dam breach problems with the proposed anti-diffusive central scheme.

\subsubsection{Dam break in one-dimension}
\begin{figure}
	\begin{subfigure}[b]{\textwidth}
		\centering
		\includegraphics[scale=0.6]{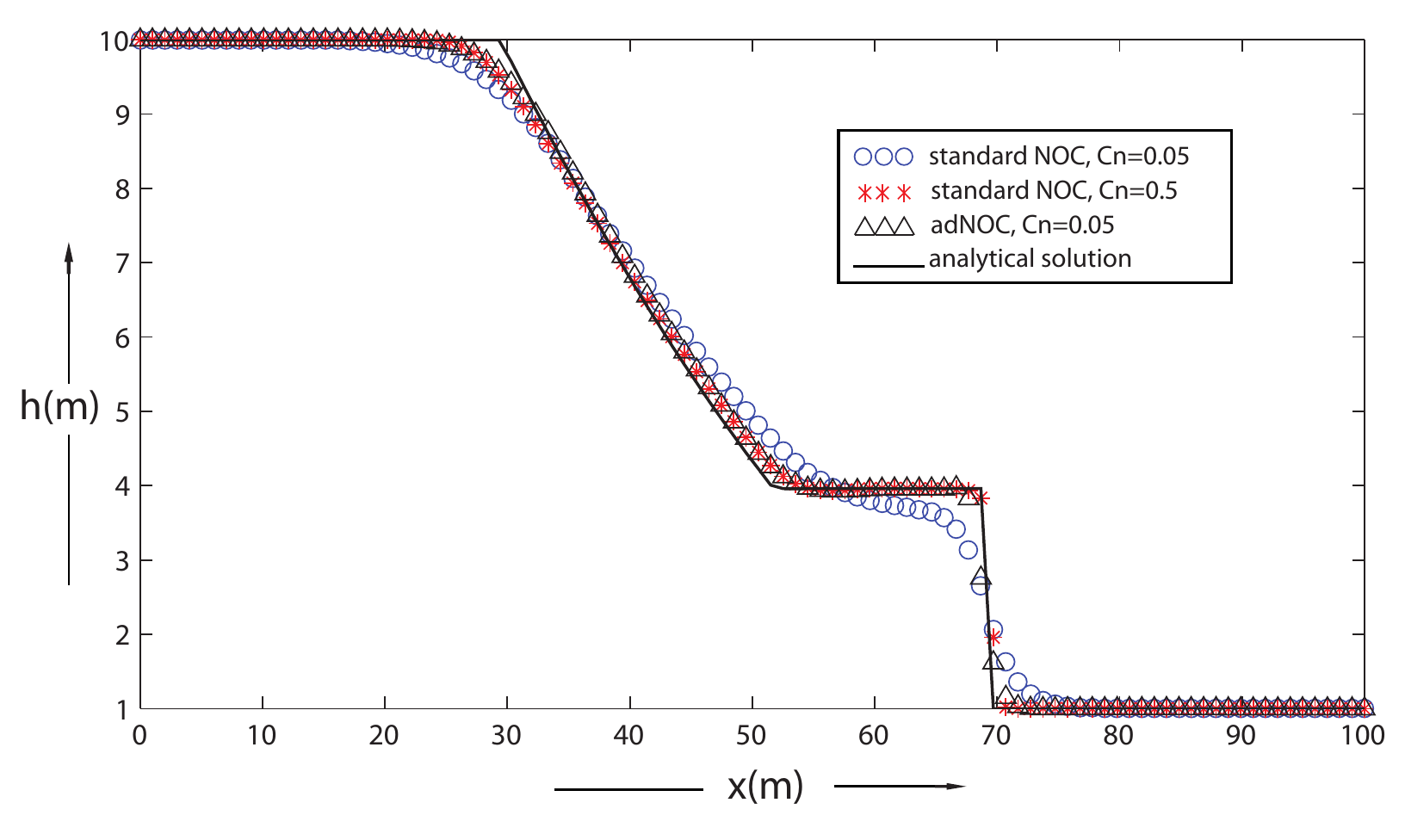}
		\caption{Water depth (\textit{h}) after two seconds of dam breach as
		calculated by different NOC schemes.} 
		\label{fig:1ddama}
	\end{subfigure}

	\begin{subfigure}[b]{\textwidth}
		\centering
		\includegraphics[scale=0.6]{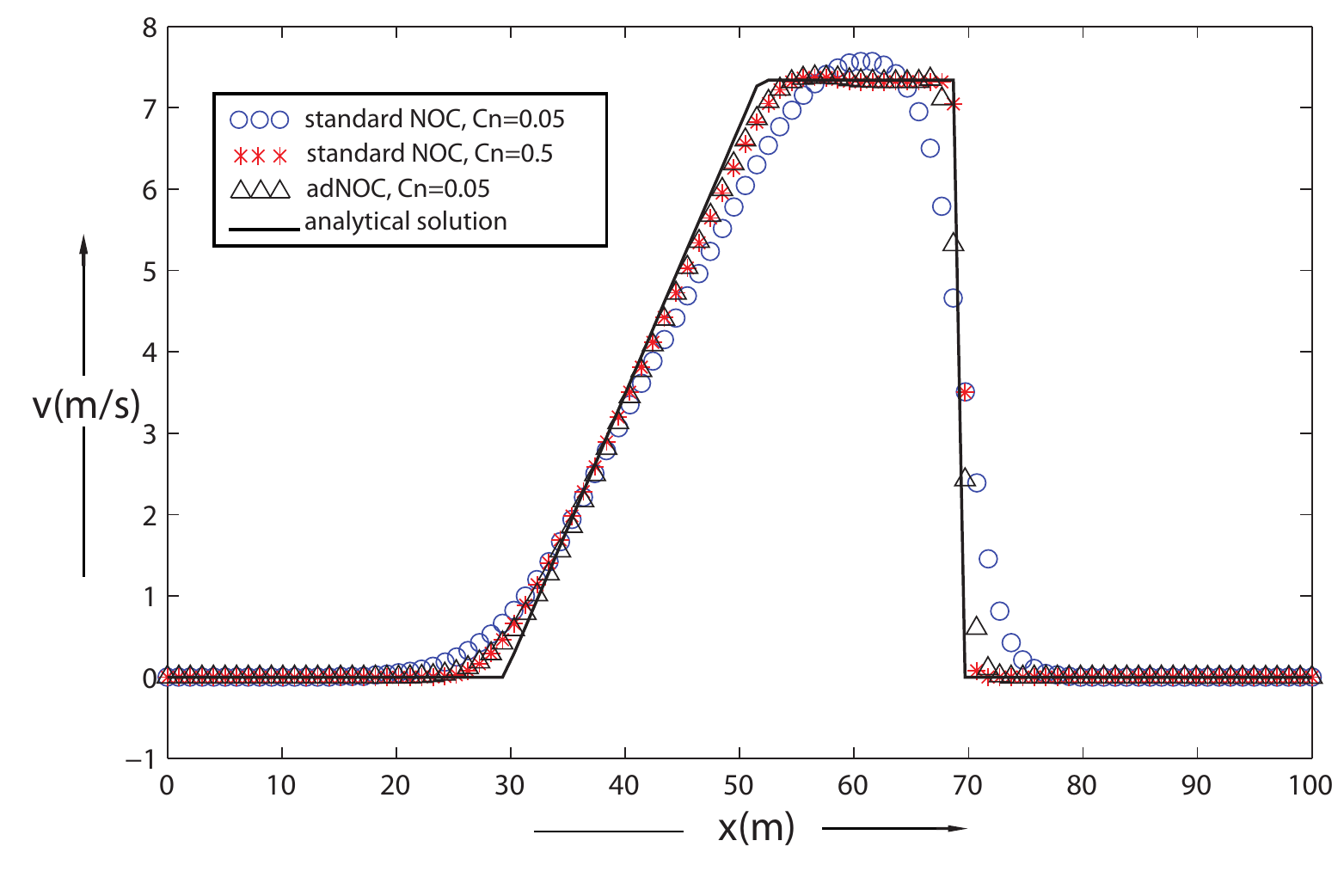}
		\caption{Velocity (\textit{v}) after two seconds of dam breach as calculated
		by different NOC schemes.} 
		\label{fig:1ddamb}
	\end{subfigure}
	\caption{Solution for 1d-dam breach experiment.}\label{fig:1ddam}
\end{figure}

The first test case consists of dam breach problem in one dimension, an
analytical solution for which is available (see \citep{toro2001shock}). A 100
meters long domain is initially separated by a dam in the middle separating two regions with water depths of 10 meters and 1 meter. The dam is removed instantaneously at time zero, which results in a sharp shock wave propagating downstream along
with a smooth rarefaction wave propagating upstream. The solution is obtained
first by the standard NOC scheme with the time steps calculated using courant
numbers of 0.5 and 0.05 using:
\begin{equation}\label{eq:27}
\Delta t=Cn\{min(\min_{i,j}\frac{\Delta x}{(\mid
u\mid+a)_{i,j}}),\min_{i,j}\frac{\Delta y}{(\mid 
v\mid+a)_{i,j}})\},
\end{equation}
where $a=\sqrt{gh}$ represents the eigenvalues of the shallow water 
equations,
see \cite{stecca2012finite}. 
The domain is discretized with 100 equally spaced cells (i.e., $\Delta x=1 m$). The results in Fig. \ref{fig:1ddama} and
\ref{fig:1ddamb} shows the water depth ($h$) and the velocity ($v$) two
seconds after  the dam breach.
The figure shows that the  classic NOC scheme is accurate when Cn = 0.5, while
it is excessively diffusive when Cn = 0.05. Also shown in Fig. \ref{fig:1ddama}
and \ref{fig:1ddamb} are the results computed  using the proposed corrected
scheme with $\varepsilon=0.85$. The results show that the proposed correction does a very good job in eliminating the diffusion present when Cn = 0.05.

\subsubsection{Breach of circular dam} 
The second test case (investigated previously by Stecca et al.
\cite{stecca2012finite}) considers the breach of a cylindrical  tank immersed in a frictionless water body initially at rest. The initial conditions are:
\[
 \begin{cases} h(x,y,0)=2.5m &\mbox{if } x^2+y^2\leq r^2 \\ 
h(x,y,0)=0.5m &\mbox{if } x^2+y^2> r^2 \\
u(x,y,0)=v(x,y,0)=0 &\mbox{} \forall x,y \end{cases} 
\]
where $r=2.5 m$ is the tank radius. The spatial domain is a 40 m x 40 m square that was  discretized with
100 cells in both $x$ and $y$ directions.  Fig. \ref{fig:circdam2d} and \ref{fig:comp} shows numerical
results  after 1.4 seconds, computed with different variants of the NOC scheme.
Also shown in Fig. \ref{fig:comp}  is a high resolution numerical reference
solution, calculated using 1000 cells in each direction using the NOC scheme
with 1000 cells in each direction and courant number of 0.5.
Results show that  the standard NOC scheme exhibits varying degrees of
numerical dispersion, depending on the time step utilized and scheme-order: 
although the second order scheme is clearly less diffusive than the first order
scheme, both schemes exhibit excessive smearing  as the courant number is
decreased. This dispersion is markedly reduced by the proposed anti-diffusion
correction ($\varepsilon$= 0.75 ),  especially when  small courant numbers are
used. These results compare  favourably with results computed for the same test
case by Stecca et al. \cite{stecca2012finite}, illustrated in Fig.
\ref{fig:stecca}.

\begin{figure}
	\begin{subfigure}{\textwidth}
		\centering
		\includegraphics[scale=0.7]{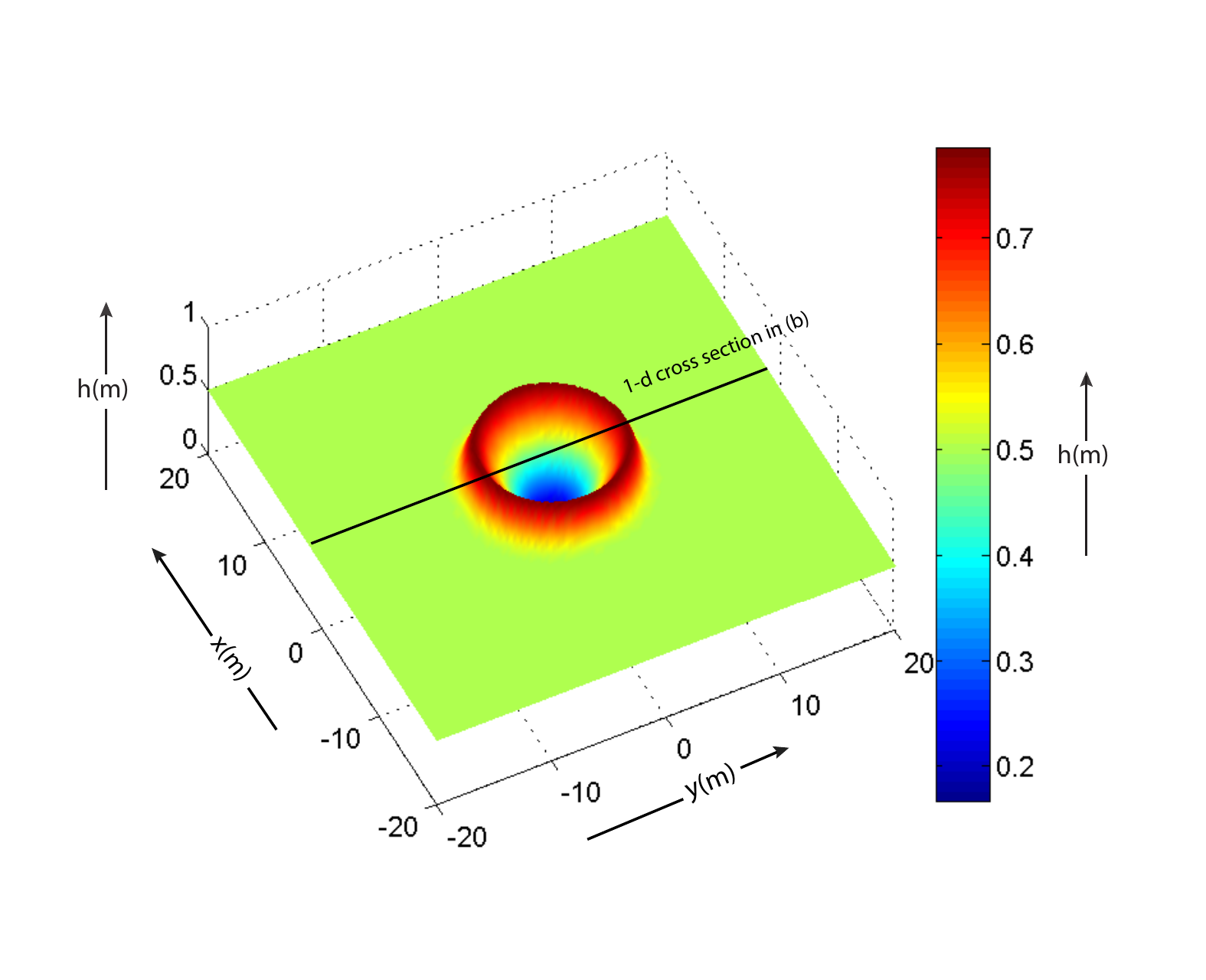}
		\caption{Water depth after 1.4 seconds of the dam break as 
		calculated by anti-diffusion first order central scheme.}
		\label{fig:circdam2d}
	\end{subfigure}

	\begin{subfigure}{\textwidth}
		\centering
		\includegraphics[scale=0.5]{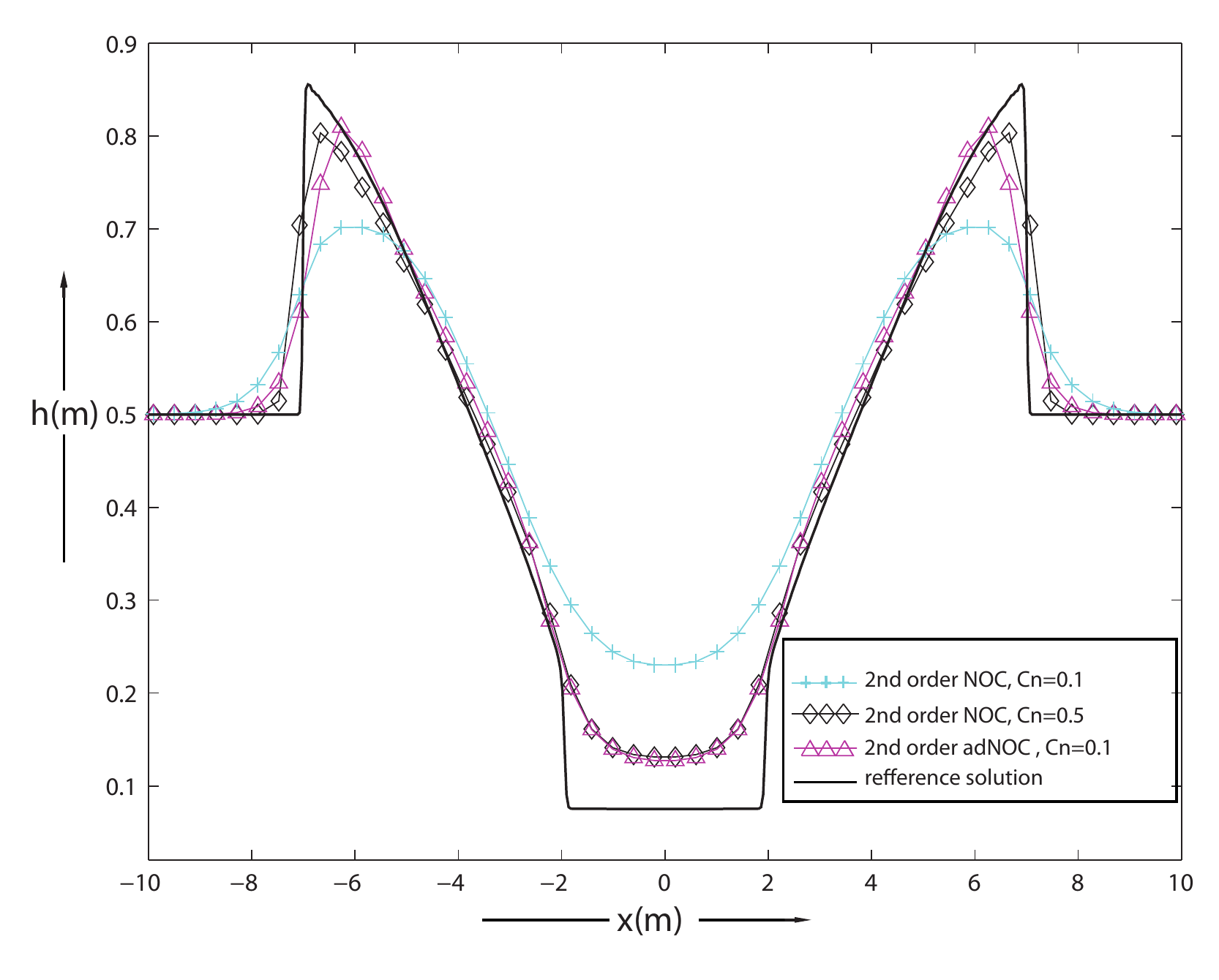}
		\caption{The one dimensional cross section of water depths after 
		1.4 seconds as calculated by different schemes.}
		\label{fig:comp}
	\end{subfigure}
	\caption{Solution for circular dam breach experiment.}\label{fig:2ddam}
\end{figure}

\begin{figure}
	\ContinuedFloat
	\begin{subfigure}{\textwidth}
	\centering
	\includegraphics[scale=0.6]{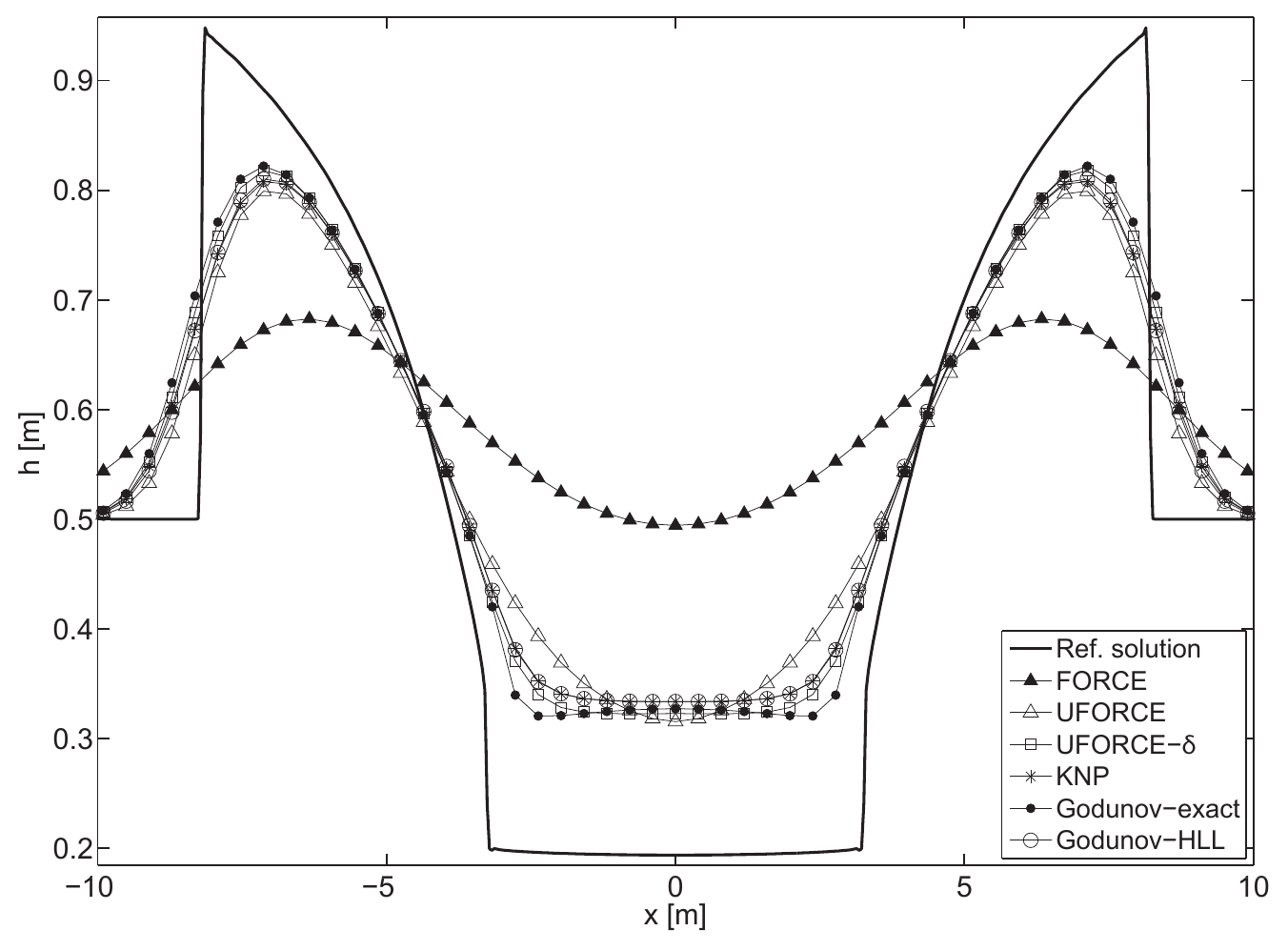}
	\caption{Comparison of numerical solutions calculated for the circular dam
		break test case, performed by \cite{stecca2012finite}. Graph shows water
		height as a function of distance after 1.4 seconds. The Courant number used
		was 0.1. The compared schemes are FORCE (a first order central differencing
		method), UFORCE and KNP  (both of which are central-upwind method) and Godunov (using
		either an exact Riemann solver or the HLL approximate Riemann solver).}
	\label{fig:stecca}
	\end{subfigure}
	\caption{Solution for circular dam breach experiment.}\label{fig:2ddamc}
\end{figure}

\subsection{Flow over mobile bed}
This test case illustrates the problem faced by central schemes for
coupled systems and it highlights how the anti-diffusive correction presented in 
this paper largely 
eliminates this short coming. The test case has been investigated in at least two earlier 
studies
(\cite{vcrnjaric2004extension,caleffi2007high}) using high 
order ENO and
CWENO schemes.
The test case considers one dimensional  flow of water over a mobile bed,
resulting in sediment transport and bed deformation in the direction of flow.
This problem is modelled by coupling the shallow water equations to the  Exner
equation (see \cite{paola2005generalized}) for evolution of the 
bed profile. The complete
system in one dimension is given by:
\[
\frac{\partial U}{\partial t}+\frac{\partial F}{\partial x}=S,
\]
where the conserved variable vector $U$, the flux vector $F$ and 
the source
vector $S$ are given by:
\[U=
\left[
\begin{array}{c}
h\\
hv\\
z\\
\end{array}
\right],
\]
\[
F=
\left[
\begin{array}{c}
hv\\
hv^2 + \frac{1}{2}gh^2\\
\frac{1}{1-\phi}q_z\\
\end{array}
\right],
\]
\[
S=
\left[
\begin{array}{c}
0\\
-gh\frac{\partial z}{\partial x}\\
0\\
\end{array}
\right].
\]
In these equations $h$ is the water depth, $v$ is the flow velocity, $g$ is the
acceleration due to gravity, $z$ is the bed elevation, $\phi$ is the substrate
porosity and $q_z$ is the sediment flux, assumed here to be given by the Grass
law \cite{grass1981sediment}:
\[
q_{z}(v)=Av\mid v\mid^{m-1},
\]
where   $A$ 
depends on the
type of sediment and signifies the coupling of the sediment with 
the water flow
and $m$ is the exponent in the range  $1\leq m\leq 4$. 

\begin{figure}
\centering
\includegraphics[scale=0.6]{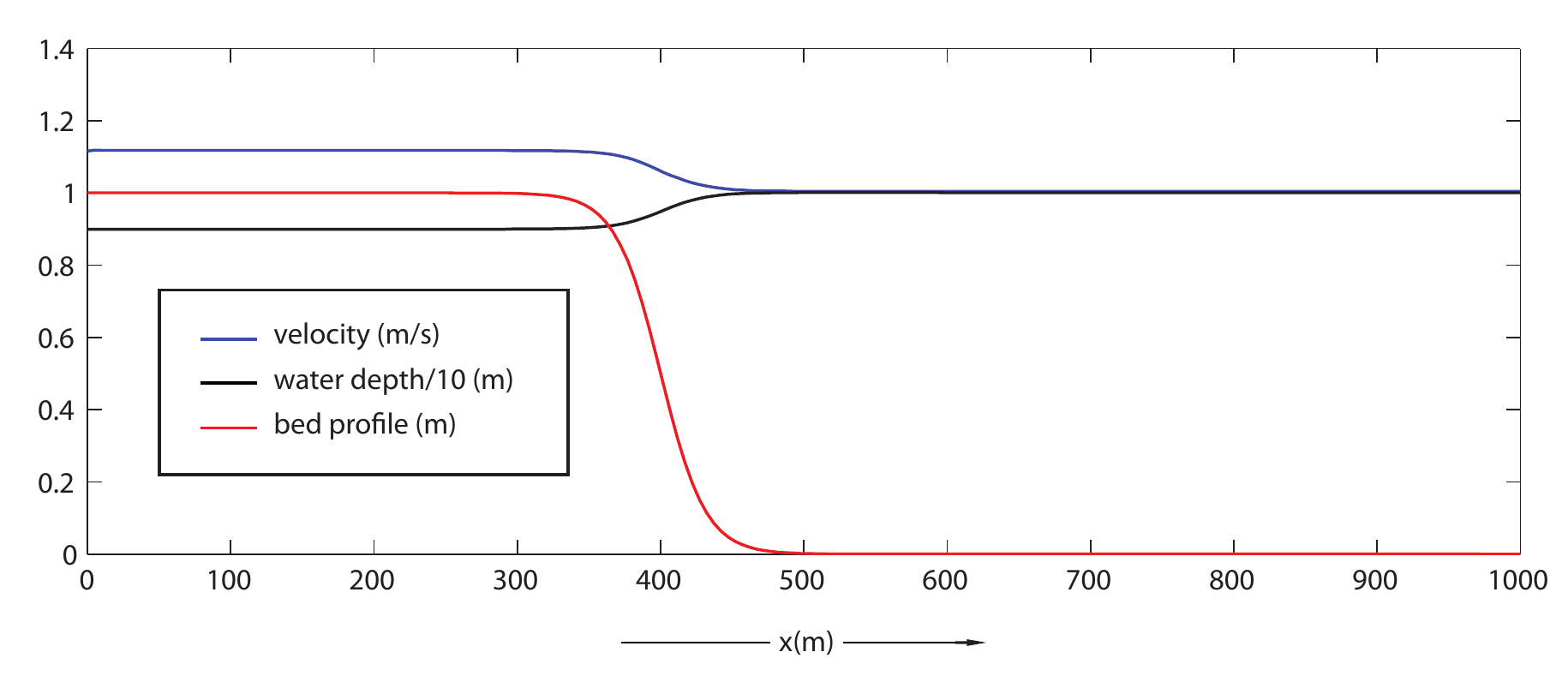}

\caption{Initial conditions for the flow over mobile bed test 
case. The water
depth is divided by 10 for better viewing.}
\label{fig:initial_conditions}
\end{figure}

\begin{figure}
\centering
\includegraphics[scale=0.5]{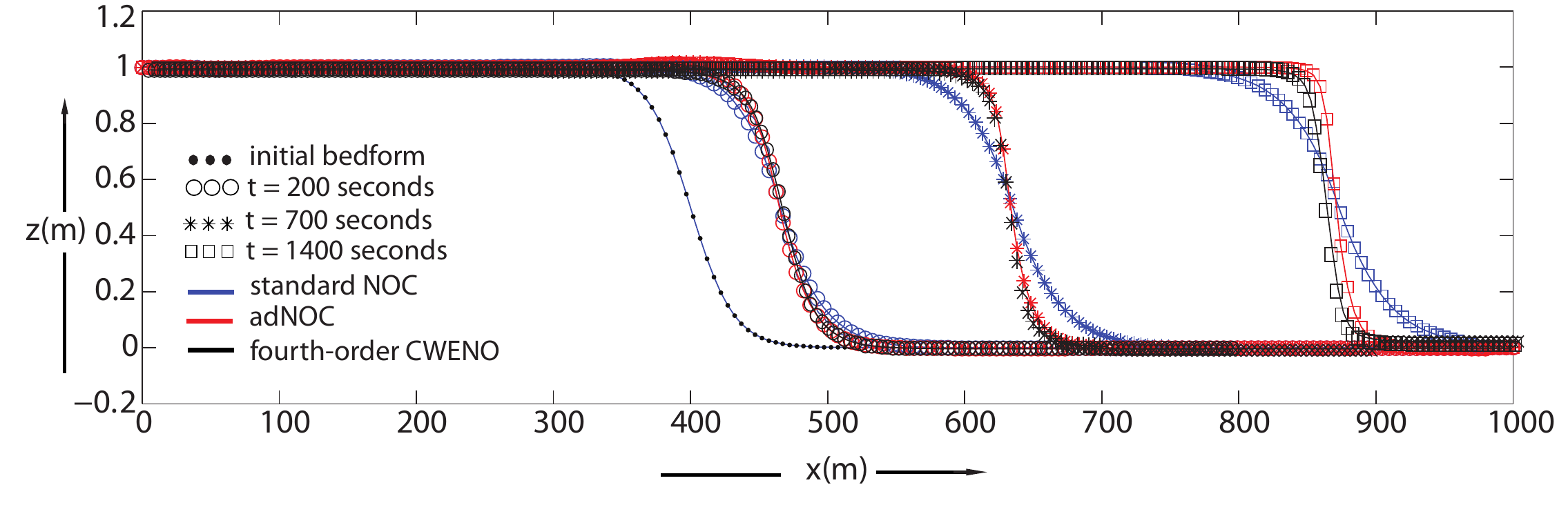}

\caption{The bed-form after 200, 700 and 1400 seconds of sediment 
transport as
calculated by standard NOC, anti-diffusive NOC and fourth order CWENO 
schemes.}
\label{fig:bedform}
\end{figure}

The test consists of a 1000 meters long domain with the initial 
bed
profile given by:
\[
z(x,0)=\frac{1}{1+e^{(\frac{x-400}{5\pi})}}.
\]
 The initial hydraulic 
conditions were computed assuming  a constant water level of 10 meters, a water
discharge of 10 m/s  at both upstream and downstream boundaries and a initially fixed bed,
see Fig. \ref{fig:initial_conditions}. Once a steady-state was reached (time=0),
the bed and flow conditions were tracked through time.
\begin{doublespace}
\end{doublespace}
Simulations were performed with the following parameter values:  $A=1$, $m=3$,
$\phi=0.2$ and  $\Delta x=5 m$.  The value for $A$ is notably high,  implying 
strong coupling between the flow and underlying substrate. Values for A  in
nature are often ca. 0.001, implying much weaker coupling, which would further
aggravate numerical dispersion with numerical methods. Calculations with the
anti-diffusive scheme were performed using $\varepsilon=1$ for the Exner
equation and $\varepsilon=0.92$ for the shallow water equations. The courant
number used to calculate the time step is 0.45 for standard NOC scheme and 0.2
for anti-diffusive NOC scheme. Fig. \ref{fig:bedform}  shows the computed bed
profile after 200, 700 and 1400 seconds. It can be seen that the standard NOC
scheme suffers from excessive diffusivity while the anti-diffusive NOC schemes
performs on par with the fourth order CWENO scheme
presented by \cite{caleffi2007high}.

\begin{doublespace}
\end{doublespace}

\section{Conclusion}
\label{sec:Conclusion}
The excessive diffusion suffered by explicit non-oscillatory central
differencing (NOC) schemes when utilizing small time
steps is tackled. An anti-diffusive version of the well-known Nessyahu-Tadmor central scheme
is presented and tested.  The condition for stability is derived and the stable
value of courant number with respect to other parameters is shown.
The proposed scheme is validated using a number of test cases involving the shallow water equations  in one and two dimensions. 
The corrected scheme is shown to significantly improve the numerical dispersion
exhibited by the standard NOC scheme, especially when small time steps are used.

\singlespacing
\bibliography{biblio}

\begin{thebibliography}{36}
\providecommand{\natexlab}[1]{#1}
\providecommand{\url}[1]{\texttt{#1}}
\expandafter\ifx\csname urlstyle\endcsname\relax
  \providecommand{\doi}[1]{doi: #1}\else
  \providecommand{\doi}{doi: \begingroup \urlstyle{rm}\Url}\fi

\bibitem[Abreu et~al.(2009)Abreu, Pereira, and Ribeiro]{abreu2009central}
E~Abreu, F~Pereira, and S~Ribeiro.
\newblock Central schemes for porous media flows.
\newblock \emph{Computational \& Applied Mathematics}, 28\penalty0
  (1):\penalty0 87--110, 2009.

\bibitem[Anile et~al.(2001)Anile, Nikiforakis, and
  Pidatella]{anile2001assessment}
A~Marcello Anile, Nikolaos Nikiforakis, and Rosa~M Pidatella.
\newblock Assessment of a high resolution centered scheme for the solution of
  hydrodynamical semiconductor equations.
\newblock \emph{SIAM Journal on Scientific Computing}, 22\penalty0
  (5):\penalty0 1533--1548, 2001.

\bibitem[Arminjon et~al.(1995)Arminjon, Stanescu, and Viallon]{arminjon1995two}
P~Arminjon, D~Stanescu, and MC~Viallon.
\newblock A two-dimensional finite volume extension of the lax-friedrichs and
  nessyahu-tadmor schemes for compressible flows.
\newblock In \emph{Proceedings of the 6th Int. Symposium on Comp. Fluid
  Dynamics}, pages 7--14, 1995.

\bibitem[Balb{\'a}s and Qian(2009)]{balbas2009non}
Jorge Balb{\'a}s and Xin Qian.
\newblock Non-oscillatory central scheme for 3d hyperbolic conservation laws.
\newblock In \emph{Proc. Sympos. Appl. Math}, volume~67, pages 389--398, 2009.

\bibitem[Balb{\'a}s et~al.(2004)Balb{\'a}s, Tadmor, and Wu]{balbas2004non}
Jorge Balb{\'a}s, Eitan Tadmor, and Cheng-Chin Wu.
\newblock Non-oscillatory central schemes for one-and two-dimensional mhd
  equations: I.
\newblock \emph{Journal of Computational Physics}, 201\penalty0 (1):\penalty0
  261--285, 2004.

\bibitem[Bianco et~al.(1999)Bianco, Puppo, and Russo]{bianco1999high}
Franca Bianco, Gabriella Puppo, and Giovanni Russo.
\newblock High-order central schemes for hyperbolic systems of conservation
  laws.
\newblock \emph{SIAM Journal on Scientific Computing}, 21\penalty0
  (1):\penalty0 294--322, 1999.

\bibitem[Bryson et~al.(2005)Bryson, Kosovichev, and Levy]{bryson2005high}
Steve Bryson, Alexander Kosovichev, and Doron Levy.
\newblock High-order shock-capturing methods for modeling dynamics of the solar
  atmosphere.
\newblock \emph{Physica D: Nonlinear Phenomena}, 201\penalty0 (1):\penalty0
  1--26, 2005.

\bibitem[Caleffi et~al.(2007)Caleffi, Valiani, and Bernini]{caleffi2007high}
Valerio Caleffi, Alessandro Valiani, and Anna Bernini.
\newblock High-order balanced cweno scheme for movable bed shallow water
  equations.
\newblock \emph{Advances in water resources}, 30\penalty0 (4):\penalty0
  730--741, 2007.

\bibitem[Canestrelli and Toro(2012)]{canestrelli2012restoration}
Alberto Canestrelli and Eleuterio~F Toro.
\newblock Restoration of the contact surface in force-type centred schemes i:
  Homogeneous two-dimensional shallow water equations.
\newblock \emph{Advances in Water Resources}, 47:\penalty0 88--99, 2012.

\bibitem[Chertock et~al.(2014)Chertock, Kurganov, and Liu]{chertock2014central}
Alina Chertock, Alexander Kurganov, and Yu~Liu.
\newblock Central-upwind schemes for the system of shallow water equations with
  horizontal temperature gradients.
\newblock \emph{Numerische Mathematik}, 127\penalty0 (4):\penalty0 595--639,
  2014.

\bibitem[{\v{C}}rnjari{\'c}-{\v{Z}}ic et~al.(2004){\v{C}}rnjari{\'c}-{\v{Z}}ic,
  Vukovi{\'c}, and Sopta]{vcrnjaric2004extension}
Nelida {\v{C}}rnjari{\'c}-{\v{Z}}ic, Senka Vukovi{\'c}, and Luka Sopta.
\newblock Extension of eno and weno schemes to one-dimensional sediment
  transport equations.
\newblock \emph{Computers \& fluids}, 33\penalty0 (1):\penalty0 31--56, 2004.

\bibitem[Engquist and Runborg(1999)]{engquist1999multiphase}
Bj{\"o}rn Engquist and Olof Runborg.
\newblock \emph{Multiphase computations in geometrical optics}.
\newblock Springer, 1999.

\bibitem[Godunov(1959)]{godunov1959difference}
Sergei~Konstantinovich Godunov.
\newblock A difference method for numerical calculation of discontinuous
  solutions of the equations of hydrodynamics.
\newblock \emph{Matematicheskii Sbornik}, 89\penalty0 (3):\penalty0 271--306,
  1959.

\bibitem[Grass(1981)]{grass1981sediment}
AJ~Grass.
\newblock \emph{Sediment transport by waves and currents}.
\newblock University College, London, Dept. of Civil Engineering, 1981.

\bibitem[Harten(1983)]{harten1983high}
Ami Harten.
\newblock High resolution schemes for hyperbolic conservation laws.
\newblock \emph{Journal of computational physics}, 49\penalty0 (3):\penalty0
  357--393, 1983.

\bibitem[Huynh(1995)]{huynh1995piecewise}
HT~Huynh.
\newblock A piecewise-parabolic dual-mesh method for the euler equations.
\newblock In \emph{12th AIAA Computational Fluid Dynamics Conference}, pages
  1054--66, 1995.

\bibitem[Huynh(2003)]{huynh2003analysis}
HT~Huynh.
\newblock Analysis and improvement of upwind and centered schemes on
  quadrilateral and triangular meshes.
\newblock \emph{AIAA Paper}, 3541:\penalty0 23--26, 2003.

\bibitem[Jiang and Tadmor(1998)]{jiang1998nonoscillatory}
Guang-Shan Jiang and Eitan Tadmor.
\newblock Nonoscillatory central schemes for multidimensional hyperbolic
  conservation laws.
\newblock \emph{SIAM Journal on Scientific Computing}, 19\penalty0
  (6):\penalty0 1892--1917, 1998.

\bibitem[Katsaounis and Levy(1999)]{katsaounis1999modified}
Theodoros Katsaounis and Doron Levy.
\newblock A modified structured central scheme for 2d hyperbolic conservation
  laws.
\newblock \emph{Applied mathematics letters}, 12\penalty0 (6):\penalty0 89--96,
  1999.

\bibitem[Kurganov(2002)]{kurganov2002central}
Alexander Kurganov.
\newblock Central-upwind schemes for balance laws. application to the broadwell
  model.
\newblock In \emph{Proceedings of the Third International Symposium on Finite
  Volumes for Complex Applications}, 2002.

\bibitem[Kurganov and Lin(2007)]{kurganov2007reduction}
Alexander Kurganov and Chi-Tien Lin.
\newblock On the reduction of numerical dissipation in central-upwind schemes.
\newblock \emph{Commun. Comput. Phys}, 2\penalty0 (1):\penalty0 141--163, 2007.

\bibitem[Kurganov and Pollack(2011)]{kurganov2011semi}
Alexander Kurganov and Michael Pollack.
\newblock Semi-discrete central-upwind schemes for elasticity in heterogeneous
  media.
\newblock \emph{SIAM J. Sci. Comput.(Submitted for publication)}, 2011.

\bibitem[Kurganov and Tadmor(2000)]{kurganov2000new}
Alexander Kurganov and Eitan Tadmor.
\newblock New high-resolution central schemes for nonlinear conservation laws
  and convection--diffusion equations.
\newblock \emph{Journal of Computational Physics}, 160\penalty0 (1):\penalty0
  241--282, 2000.

\bibitem[LeVeque and Le~Veque(1992)]{leveque1992numerical}
Randall~J LeVeque and Randall~J Le~Veque.
\newblock \emph{Numerical methods for conservation laws}, volume 132.
\newblock Springer, 1992.

\bibitem[Levy et~al.(2002)Levy, Puppo, and Russo]{levy2002fourth}
Doron Levy, Gabriella Puppo, and Giovanni Russo.
\newblock A fourth-order central weno scheme for multidimensional hyperbolic
  systems of conservation laws.
\newblock \emph{SIAM Journal on scientific computing}, 24\penalty0
  (2):\penalty0 480--506, 2002.

\bibitem[Liu and Tadmor(1998)]{liu1998third}
Xu-Dong Liu and Eitan Tadmor.
\newblock Third order nonoscillatory central scheme for hyperbolic conservation
  laws.
\newblock \emph{Numerische Mathematik}, 79\penalty0 (3):\penalty0 397--425,
  1998.

\bibitem[Nessyahu and Tadmor(1990)]{NessyahuTadmor1990}
H.~Nessyahu and E.~Tadmor.
\newblock Non-oscillatory central differencing schemes for hyperbolic
  conservation laws.
\newblock \emph{J. Comp. Phys.}, 1990.
\newblock \doi{10.1016/0021-9991(90)90260-8}.

\bibitem[Paola and Voller(2005)]{paola2005generalized}
C~Paola and VR~Voller.
\newblock A generalized exner equation for sediment mass balance.
\newblock \emph{Journal of Geophysical Research: Earth Surface (2003--2012)},
  110\penalty0 (F4), 2005.

\bibitem[Pudasaini and Hutter(2007)]{pudasaini2007avalanche}
Shiva~P Pudasaini and Kolumban Hutter.
\newblock \emph{Avalanche dynamics: dynamics of rapid flows of dense granular
  avalanches}.
\newblock Springer Science \& Business Media, 2007.

\bibitem[Qiu and Shu(2002)]{qiu2002construction}
Jianxian Qiu and Chi-Wang Shu.
\newblock On the construction, comparison, and local characteristic
  decomposition for high-order central weno schemes.
\newblock \emph{Journal of Computational Physics}, 183\penalty0 (1):\penalty0
  187--209, 2002.

\bibitem[Siviglia et~al.(2013)Siviglia, Stecca, Vanzo, Zolezzi, Toro, and
  Tubino]{siviglia2013numerical}
Annunziato Siviglia, Guglielmo Stecca, Davide Vanzo, Guido Zolezzi, Eleuterio~F
  Toro, and Marco Tubino.
\newblock Numerical modelling of two-dimensional morphodynamics with
  applications to river bars and bifurcations.
\newblock \emph{Advances in Water Resources}, 52:\penalty0 243--260, 2013.

\bibitem[Stecca et~al.(2012)Stecca, Siviglia, and Toro]{stecca2012finite}
Guglielmo Stecca, Annunziato Siviglia, and Eleuterio~F Toro.
\newblock A finite volume upwind-biased centred scheme for hyperbolic systems
  of conservation laws. applications to shallow water equations.
\newblock \emph{Commun Comput Phys}, 12\penalty0 (4):\penalty0 1183--1214,
  2012.

\bibitem[Sweby(1984)]{sweby1984high}
Peter~K Sweby.
\newblock High resolution schemes using flux limiters for hyperbolic
  conservation laws.
\newblock \emph{SIAM journal on numerical analysis}, 21\penalty0 (5):\penalty0
  995--1011, 1984.

\bibitem[Tai et~al.(2002)Tai, Noelle, Gray, and Hutter]{tai2002shock}
Yih-Chin Tai, S~Noelle, JMNT Gray, and Kolumban Hutter.
\newblock Shock-capturing and front-tracking methods for granular avalanches.
\newblock \emph{Journal of Computational Physics}, 175\penalty0 (1):\penalty0
  269--301, 2002.

\bibitem[Toro(2001)]{toro2001shock}
Eleuterio~F Toro.
\newblock \emph{Shock-capturing methods for free-surface shallow flows}.
\newblock Wiley, 2001.

\bibitem[Toro(2009)]{toro2009riemann}
Eleuterio~F Toro.
\newblock \emph{Riemann solvers and numerical methods for fluid dynamics: a
  practical introduction}.
\newblock Springer Science \& Business Media, 2009.

\end{thebibliography}

\clearpage
\begin{appendices}

\section{Anti-diffusive central scheme in 2-dimensions}
\label{app:anti-diffusive central scheme in 2-dimensions}
A two-dimensional extension of the Nessyahu-Tadmor scheme can be seen in
\cite{jiang1998nonoscillatory,pudasaini2007avalanche}. Following the discussion
in section 4 for anti-diffusive central scheme in one dimension (Eq.
\ref{eq:23}), along with discussion in the above mentioned references, the
anti-diffusive central scheme in two dimensions is given by:
\[
\overline{u}_{p+1/2,q+1/2}^{n+1}=\frac{1}{4}\{\hat{u}_{p,q}^{n}+\hat{u}_{p+1,q}^{n}+\hat{u}_{p,q+1}^{n}+\hat{u} _{p+1,q+1}^{n}\} \]

\[ +\frac{\Delta x}{16}(1-\varepsilon)\{\sigma_{p,q}^{x}-\sigma_{p+1,q}^{x}- \sigma_{p+1,q+1}^{x}+\sigma_{p,q+1}^{x}\} \]

\[ +\frac{\Delta y}{16}(1-\varepsilon)\{\sigma_{p,q}^{y}+\sigma_{p+1,q}^{y}- \sigma_{p+1,q+1}^{y}-\sigma_{p,q+1}^{y}\}-\varepsilon \Psi \]

\[ -\frac{\Delta t}{2\Delta x}\{f(\overline{u}_{p+1,q}^{n+1/2})+f(\overline{u}_{p+1,q+1}^{n+1/2}) -f(\overline{u}_{p,q}^{n+1/2})-f(\overline{u}_{p,q+1}^{n+1/2})\} \]

\[ -\frac{\Delta t}{2\Delta y}\{g(\overline{u}_{p,q+1}^{n+1/2})+g(\overline{u}_{p+1,q+1}^{n+1/2}) -g(\overline{u}_{p,q}^{n+1/2})-g(\overline{u}_{p+1,q}^{n+1/2})\} \]

\[ +\frac{\Delta t}{4}\{s(\overline{u}_{p+1/4,q+1/4}^{n+1/2})+s(\overline{u}_{p+3/4,q+1/4 }^{n+1/2})+s(\overline{u}_{p+3/4,q+3/4}^{n+1/2})+s(\overline{u}_{p +1/4,q+3/4}^{n+1/2})\}. \]

Where $\Psi$ is the 2D anti-diffusive component of the slopes given by:
\[
\Psi=-\frac{3}{4}\overline{u}_{p+1/2,q+1/2}^{n-1}
\]
\[
+\frac{1}{8}
(\overline{u}_{p+1/2,q-1/2}^{n-1}+\overline{u}_{p+1/2,q+3/2}^{n-
1}+\overline{u}_{p-1/2,q+1/2}^{n-1}+\overline{u}_{p+3/2,q+1/2}^{n- 1}) \]
\begin{equation}
+\frac{1}{16}(\overline{u}_{p-1/2,q-1/2}^{n-
1}+\overline{u}_{p+3/2,q-1/2}^{n-1}+\overline{u}_{p-1/2,q+3/2}^{n-
1}+\overline{u}_{p+3/2,q+3/2}^{n-1})
\end{equation}
\end{appendices}

\end{document}